\documentclass[journal,fleqn]{IEEEtran}
\usepackage{amsmath,amsfonts}
\usepackage{url}
\usepackage{cite}
\usepackage{lipsum}
\usepackage{bm}
\usepackage[colorlinks=true,
            linkcolor=blue,      
            citecolor=blue,      
            urlcolor=blue]{hyperref}
\usepackage{multirow}
\usepackage{graphicx}
\usepackage{tabularx}
\usepackage{booktabs}
\usepackage{array}
\newcolumntype{Y}{>{\centering\arraybackslash}X}

\newcommand{\mt}[1]{\textit{#1}} 
\usepackage{enumitem}
\newlist{Indices}{description}{1}
\setlist[Indices]{style=nextline, leftmargin=10mm, labelindent=0cm, itemsep=0.2mm, font=\normalfont}

\newlist{Parameters}{description}{1}
\setlist[Parameters]{style=nextline, leftmargin=18mm, labelindent=0cm, itemsep=0.2mm, font=\normalfont}

\newlist{Variables}{description}{1}
\setlist[Variables]{style=nextline, leftmargin=15mm, labelindent=0cm, itemsep=0.2mm, font=\normalfont}

\begin{document}

\title{Neural Two-Stage Stochastic Optimization for Solving Unit Commitment Problem}
\author{Zhentong Shao, \textit{Member, IEEE}, Jingtao Qin, \textit{Graduate Student Member, IEEE}, Nanpeng Yu, \textit{Senior Member, IEEE}

\thanks{Z. Shao, J. Qin and N. Yu are with the Department of Electrical and Computer Engineering, University of California, Riverside, CA 92521 USA (e-mail: zhentons@ucr.edu; jqin020@ucr.edu; nyu@ece.ucr.edu).
}

\vspace{-9mm}
}

\maketitle

\begin{abstract}
    This paper proposes a neural stochastic optimization method for efficiently solving the two-stage stochastic unit commitment (2S-SUC) problem under high-dimensional uncertainty scenarios. The proposed method approximates the second-stage recourse problem using a deep neural network trained to map commitment decisions and uncertainty features to recourse costs. The trained network is subsequently embedded into the first-stage UC problem as a mixed-integer linear program (MILP), allowing for explicit enforcement of operational constraints while preserving the key uncertainty characteristics. A scenario-embedding network is employed to enable dimensionality reduction and feature aggregation across arbitrary scenario sets, serving as a data-driven scenario reduction mechanism. 
    Numerical experiments on IEEE 5-bus, 30-bus, and 118-bus systems demonstrate that the proposed neural two-stage stochastic optimization method achieves solutions with an optimality gap of less than 1\%, while enabling orders-of-magnitude speedup compared to conventional MILP solvers and decomposition-based methods. Moreover, the model’s size remains constant regardless of the number of scenarios, offering significant scalability for large-scale stochastic unit commitment problems.
\end{abstract}

\vspace{-1mm}

\section*{Nomenclature}
\subsection{Indices}
\begin{Indices}
    \item[$d$] Index of system demands, $d \in \mathcal{N}_{d}$.
    \item[$g$] Index of generator units, $g \in \mathcal{N}_{g}$.
    \item[$l$] Index of transmission lines, $l \in \mathcal{N}_{l}$.
    \item[$i,j$] Index of hidden units of neural network.
    \item[$m$] Index of layers of neural network , $m \in \{1,\ldots,\ell\}$.
    \item[$n$] Index of system nodes, $n \in \mathcal{N}_{n}$.
    \item[$s$] Index of uncertainty scenarios, $s \in \mathcal{S}$.
    \item[$t$] Index of time periods, $t \in \mathcal{N}_{t}$.
\end{Indices}

\subsection{Parameters}
\begin{Parameters}
    \item[$\mathbf{b}^{m}$] Bias vector of $m$-th neural network layer, with its $i$-th element denoted as $b_{i}^{m}$.
    \item[$C_{g}(\cdot)$] Function of fuel cost of unit $g$.
    \item[$C_{g}^{\mt{SU}}, C_{g}^{\mt{SD}}$] Start-up and shut-down cost of unit $g$.
    \item[$\epsilon$] Parameter controlling the perturbation level in sampling, with $\epsilon \in (0, 1)$.
    \item[$\eta$] Parameter controlling the restriction of the solution space, with $\eta \in (0, 1)$.
    \item[$F(\cdot)$] Rescaling function used to rescale the neural network output to the original cost domain.
    \item[$\overline{F}_{l}$] Transmission capacity of line $l$.
    \item[$\Gamma_{l,n}$] Power transfer distribution factors (PTDF) of node $n$ with respect to line $l$.
    \item[$\gamma_{g,t}$] Randomly generated coefficients for data sampling, which is sampled uniformly from the interval $[-1, 1]$.
    \item[$\mathbf{h}^{m}$] Hidden states of the $m$-th neural network layer.
    \item[$J$] Expected operational cost of the second-stage recourse problem.
    \item[$\underline{P}_g, \overline{P}_g$] Minimum and maximum power output for unit $g$.
    \item[$\pi_{s}$] Probability of scenario $s$ being realized.
    \item[$\Phi_{\mt{Main}}$] ReLU-activated fully connected neural network embedded in the optimization model.
    \item[$\Phi_{\mt{Scn}}$] Scenario-embedding neural network for uncertainty feature extraction.
    \item[$Q(\cdot,\cdot)$] Abstract function representing the second-stage recourse problem.
    \item[$R_g^{\mt{UP}}, R_g^{\mt{DN}}$] Ramp up and down limits for unit $g$.
    \item[$T_g^{\mt{UP}}, T_g^{\mt{DN}}$] Minimum up and down time limits for unit $g$.
    \item[$\mathbf{W}^{m}$] Weight matrix of the $m$-th layer in the neural network, where $w_{ij}^{m}$ denotes the element in the $i$-th row and $j$-th column.
    \item[$\bm{\xi}_{s}$] Vector of uncertain net loads in scenario $s$.
    \item[$\bm{\zeta}$] Vector of scenario embedding features produced by the scenario-embedding network.
\end{Parameters}

\subsection{Variables}
\begin{Variables}
    \item[$p_{d,t}$] Net load $d$ at time $t$.
    \item[$p_{g,t}$] Power generation of unit $g$ at time $t$.
    \item[$\psi$] Prediction of the neural network.
    \item[$u_{g,t}$] Start-up status of unit $g$ at time $t$, $u_{g,t} \in \{0,1\}$.
    \item[$v_{g,t}$] Shut-down status of unit $g$ at time $t$, $v_{g,t} \in \{0,1\}$.
    \item[$z_{g,t}$] Commitment status of unit $g$ at time $t$, $z_{g,t} \in \{0,1\}$.
    \item[$z_{i}^{m}$] Binary variable representing the activation status of the \texttt{ReLU} function for the $i$-th neuron in $m$-th layer.
    \item[$\bm{x}$] Vector of unit status over the scheduling horizon, $\bm{x} = \{u_{g,t},v_{g,t},z_{g,t}\}_{\forall g, \forall t}$.
    \item[$\bm{y}$] Vector of generator dispatch levels over the scheduling horizon.
    \item[$\bm{z}$] Vector of commitment status, $\bm{z} = \{z_{g,t}\}_{\forall g, \forall t}$.
\end{Variables}

\section{Introduction}
\subsection{Research Motivation}
\IEEEPARstart{W}{ith} renewable energy sources become increasingly dominant in power systems, the associated uncertainty has emerged as a critical factor affecting system reliability and operational efficiency. To address this challenge, two-stage stochastic unit commitment (2S-SUC) has become a widely adopted framework for managing day-ahead system operations under uncertainty \cite{wu2007stochastic}. The extensive form is commonly employed to solve the 2S-SUC, where uncertainty is captured through a finite set of scenarios sampled from a given probability distribution \cite{entensive-form}. This leads to a computationally intensive optimization problem whose size scales linearly with the number of introduced scenarios. Consequently, the practical application of 2S-SUC remains limited by its computational complexity, which arises primarily from the trade-off between accurately representing uncertainty and maintaining scalability when solving problems with large scenario sets. 

To address the computational challenges of 2S-SUC, existing studies can be broadly classified into two main categories: decomposition techniques \cite{nasri2015network} and scenario reduction strategies \cite{feng2016solution}, both aiming to enhance tractability while maintaining model fidelity. In contrast to existing methods, this paper presents a novel perspective by introducing a neural two-stage stochastic optimization framework. In this method, the second stage of the 2S-SUC problem is approximated using a neural network surrogate, which is subsequently embedded into the optimization framework as a mixed-integer linear program (MILP). This formulation enables the explicit enforcement of hard constraints on first-stage commitment decisions, offering a key advantage over conventional direct-mapping techniques. The proposed approach can be viewed as a learning-based acceleration framework coupled with a data-driven scenario reduction strategy. Numerical experiments indicate that the proposed approach is over 100 times faster than conventional solvers, while maintaining a typical optimality gap within 1\%.

\vspace{-3mm}
\subsection{Literature Review}
\subsubsection{Stochastic Unit Commitment}
The stochastic unit commitment (SUC) problem is commonly modeled and solved using the scenario-based extensive form \cite{zheng2014stochastic}. A few studies have reformulated SUC as a robust optimization problem \cite{zhai2016transmission}, which often results in more conservative and complex formulations and is not the focus of this work. Within the extensive form framework, two main technical directions have been developed to accelerate the solution process: optimization-based decomposition methods and mathematical scenario reduction strategies. In most cases, these approaches are integrated and applied jointly to enhance computational efficiency. Typical optimization-based decomposition methods include Benders decomposition \cite{zhang2017chance}, column-and-constraint generation \cite{lee2025column}, and Lagrangian decomposition \cite{wu2007stochastic}. Meanwhile, representative scenario reduction strategies include data-driven identification of worst-case scenarios \cite{zhao2015data}, scenario mapping approaches \cite{du2018scenario}, and scenario classification methods \cite{blanco2017efficient,zhang2023stochastic}. In addition, a comparative evaluation of conventional scenario reduction techniques is provided in \cite{dvorkin2014comparison}.

Recent studies have proposed learning-based methods to improve SUC solutions, such as those presented in \cite{yurdakul2023predictive, chen2024towards}. These works primarily utilize learning-based approaches to identify economically significant scenarios, demonstrating improved economic performance over traditional sample average approximation techniques. Moreover, learning-based methodologies for rapid deterministic UC solutions have attracted growing attention and exhibit potential for extension to 2S-SUC problems, as exemplified by physics-informed graph neural networks (GNN) proposed in \cite{qin2025physics} and reinforcement learning-integrated surrogate Lagrangian relaxation frameworks described in \cite{zhu2025reinforcement}. Nevertheless, existing learning-based strategies typically rely on direct mappings from system demands and generation profiles to UC decisions, providing effective initial solutions to accelerate MILP solvers. However, such approaches often fail to guarantee strict compliance with operational constraints. In contrast, the proposed approach fundamentally differs from prior methods by embedding neural networks directly within the optimization framework, enabling constraint-aware decision-making during the solution process. 

This paper assumes the second-stage recourse problem of the 2S-SUC formulation to be a linear program (LP), which is a common modeling assumption in existing 2S-SUC literature. When quick-start units are considered, as in \cite{liu2010extended, zheng2013decomposition}, binary commitment decisions are introduced in the second stage, thereby transforming the recourse problem into a more complex MILP. Notably, the proposed neural stochastic optimization approach encodes the second-stage recourse problem using neural networks, enabling inherent adaptability to diverse and potentially non-convex problem structures. However, to clearly illustrate the effectiveness of the method, this paper does not include mixed-integer recourse instances in the case studies, but instead, focuses on evaluating the method's capability to handle large scenario sets within the 2S-SUC framework. 

\subsubsection{Neural Network Embedded Optimization}
The core of this paper is to embed neural networks into optimization problems to achieve a neural two-stage stochastic optimization. While we are among the first to apply this approach to power system optimization, particularly to accelerate the solution of 2S-SUC problems, the method itself has been explored in prior works. Notably, \cite{cheng2017maximum} and \cite{tjeng2017evaluating} showed that neural networks with Rectified Linear Unit (ReLU) activations can be equivalently represented as MILP formulations. \cite{fischetti2018deep} extended this technique to adversarial learning. Subsequent refinements, such as those in \cite{grimstad2019relu,anderson2020strong}, addressed issues like big-M tuning and inactive neuron treatment. The OMLT toolbox introduced in \cite{ceccon2022omlt} further enabled scalable MILP encodings, even for large-scale networks like GNNs.
More recently, this modeling technique has been adopted in various domains: process optimization \cite{misener2023formulating}, market bidding \cite{jalving2023beyond}, reinforcement learning \cite{shengren2023optimal}, and stochastic programming \cite{patel2022neur2sp}. Our work builds upon previous research and extends its core ideas to address power system operations under uncertainty.

\vspace{-2mm}
\subsection{Contributions}
\vspace{-1mm}
This paper proposes a neural two-stage stochastic optimization framework to accelerate the solution of 2S-SUC problems. Specifically, the second-stage recourse problem is approximated using a deep neural network that captures the mapping from uncertainty features and commitment decisions to recourse costs. The trained neural network is then reformulated into a MILP model and integrated into the first-stage UC formulation as part of its constraints. This integration results in a surrogate reformulation of the original 2S-SUC problem, with a model size that is independent of the number of uncertainty scenarios. Consequently, the proposed approach offers substantial computational advantages in large-scale stochastic settings.
The primary contributions of this paper are summarized as follows.
\begin{enumerate}
    \item[(i)] We propose a neural two-stage stochastic optimization approach to accelerate the solution of 2S-SUC problems involving a large number of scenarios. In this framework, the second-stage recourse problem is approximated using a deep neural network, which is subsequently encoded as a MILP problem. This formulation enables the explicit integration of a learned recourse surrogate into the first-stage decision-making process, there by ensuring compliance with hard operational constraints related to commitment decisions.

    \item[(ii)] The proposed framework achieves substantial computational speedup by decoupling model complexity from the number of uncertainty scenarios, effectively supporting large-scale stochastic optimization. Empirical results demonstrate that the proposed method preserves solution quality with an optimality gap below 1\%, while reducing computation time by up to three orders of magnitude, and achieving more than 100-fold speedups in most cases.
\end{enumerate}

The remainder of the paper is organized as follows: Section II describes the problem formulation of 2S-SUC. Section III presents the proposed neural two-stage stochastic optimization method. Section IV provides numerical studies, and Section V concludes the paper.

\vspace{-2mm}
\section{Stochastic Unit Commitment Formulation}
\subsection{Objective Function}
The objective of a UC problem is to minimize the total system operating cost over the scheduling horizon. This cost includes generation costs as well as start-up and shut-down costs associated with generator's state transitions. The objective function is formulated as follows:
\vspace{-2mm}
\begin{equation}
\min_{p_{g,t}, z_{g,t}, u_{g,t}, v_{g,t}} \sum_{t \in \mathcal{N}_{t}} \sum_{g \in \mathcal{N}_{g}} \left( C_{g}(p_{g,t}) + C^{\mt{SU}}_{g} u_{g,t} + C^{\mt{SD}}_{g} v_{g,t} \right),
\label{UC-obj}
\end{equation}
where the variable definitions are provided in the Nomenclature. The term $C_g(p_{g,t})$ denotes the fuel cost function of unit $g$, which can be represented as a linear, piecewise linear, or quadratic function.

\vspace{-2mm}
\subsection{System-Level Constraints}
System-level constraints define the operational limits that couple all generation units across the power network.

\subsubsection{Power Balance Constraints}
At each time period, the total power generated by all units must exactly match the net system demand:
\vspace{-2mm}
\begin{equation}
\sum_{g \in \mathcal{N}_{g}} p_{g,t} - \sum_{d \in \mathcal{N}_{d}} p_{d,t} = 0, \quad \forall t \in \mathcal{N}_{t}.
\label{UC-power-balance}
\end{equation}

\subsubsection{DC Transmission Capacity Constraints}
The power flow on each transmission line must remain within its rated capacity and is determined as a linear function of net nodal injections via PTDFs:
\vspace{-2mm}
\begin{multline}
    - \overline{F}_{l} \leq \sum_{n \in \mathcal{N}_{n}} \Gamma_{l,n} \left( \sum_{g \in \mathcal{N}_{g}(n)} p_{g,t} + \sum_{d \in \mathcal{N}_{d}(n)} p_{d,t}\right)\leq \overline{F}_{l}, \\
    \forall l \in \mathcal{N}_{l}, \forall t \in \mathcal{N}_{t}.
\label{UC-dc-flow}
\end{multline}

\vspace{-5mm}
\subsection{Generator-Level Constraints}
Generator-level constraints define the operational limits for each generator.
\subsubsection{Power Generation Capacity Constraints}
The power output of generator $g$ at time $t$ must lie within its minimum and maximum generation limits when the unit is committed, and must be zero otherwise:
\begin{equation}
z_{g,t} \underline{P}_g \leq p_{g,t} \leq z_{g,t} \overline{P}_g, \quad \forall g \in \mathcal{N}_g,  \forall t \in \mathcal{N}_{t}.
\label{UC-power-capacity}
\end{equation}

\subsubsection{Ramping Constraints}
The power output of each generator is subject to ramp-up and ramp-down limits between successive time intervals. Let $p_{g,0}$ denote the initial power output at the start of the horizon:
\begin{equation}
- R_g^{\mt{DN}} \leq p_{g,t} - p_{g,t-1} \leq R_g^{\mt{UP}}, \quad \forall g \in \mathcal{N}_g,  \forall t \in \mathcal{N}_{t}.
\label{UC-ramping}
\end{equation}

\subsubsection{Minimum Up and Down Time Constraints}
Generator units are required to remain online for at least $T_g^{\mt{UP}}$ consecutive time periods once committed, and offline for at least $T_g^{\mt{DN}}$ time periods once decommitted. The following constraints ensure logical consistency:
\begin{align}
    &u_{g,t} - v_{g,t} = z_{g,t} - z_{g,t-1}, \quad \forall g \in \mathcal{N}_g,\forall t \in \mathcal{N}_{t}, \label{UC-logic} \\
    &\sum_{\tau = t - T_g^{\mt{UP}} + 1}^{t} u_{g,\tau} \leq z_{g,t},  \forall g \in \mathcal{N}_g,  \forall t \in \{T_g^{\mt{UP}}, \ldots, |\mathcal{N}_{t}|\}, \label{UC-min-up-time} \\
    &\sum_{\tau = t - T_g^{\mt{DN}} + 1}^{t} v_{g,\tau} \leq 1 - z_{g,t},  \forall g \in \mathcal{N}_g,  \forall t \in \{T_g^{\mt{DN}}, \ldots, |\mathcal{N}_{t}|\}. \label{UC-min-down-time}
\end{align}

\vspace{-6mm}
\subsection{Two-Stage Stochastic UC Model}
The UC model defined in \eqref{UC-obj}–\eqref{UC-min-down-time} can be reformulated as a two-stage stochastic programming (2SP) problem by treating net power demands as uncertain parameters. Since the power outputs of renewable energy sources such as wind and solar can be represented as negative demands, this paper considers net load (i.e., demand minus renewable generation) as the source of uncertainty for modeling simplicity.

To facilitate a more compact representation, the UC model under uncertainty can be expressed in the canonical form of a 2SP model as follows:
\begin{align}
    &\min_{\bm{x}} ~ \bm{c}^{\mathsf{T}} \bm{x} + \mathbb{E}_{\bm{\xi}}[Q(\bm{x}, \bm{\xi})] 
    \\
    \text{s.t.}~ &\bm{x} \in \mathcal{X},
    \\
    &Q(\bm{x}, \bm{\xi}) = 
    \left\{
    \begin{aligned}
        \min_{\bm{y}} \quad & \bm{q}^{\mathsf{T}} \bm{y} \\
        \text{s.t.} \quad & \bm{W} \bm{y} \geq \bm{h}(\bm{\xi})
        ,\quad\bm{y} \in \mathcal{Y}(\bm{x})
    \end{aligned}
    \right\}, \label{Q-function}
\end{align}
where $\bm{x}$ denotes the first-stage decision variables (i.e., unit commitment statuses $z_{g,t},u_{g,t},v_{g,t}$), $\bm{c}^{\mathsf{T}} \bm{x}$ represents the corresponding deterministic cost (i.e., start-up and shut-down costs), and $\bm{\xi}$ is the random vector representing the uncertain net load. The function $Q(\bm{x}, \bm{\xi})$ defines the second-stage recourse problem, which resolves generator dispatch under given commitment decisions and uncertainties. 

The formulation of $Q(\bm{x}, \bm{\xi})$ is given by \eqref{Q-function}, where $\bm{y}$ represents the second-stage decision variables (i.e., dispatch levels $p_{g,t}$), and $\bm{q}$ denotes the generation cost coefficients. The matrix $\bm{W}$ and right-hand-side vector $\bm{h}(\bm{\xi})$ encode operational constraints (i.e., power balance and transmission capacity). The feasible set $\mathcal{Y}(\bm{x})$ is defined by operational bounds which includes generator ramping and capacity limits.

\subsection{Scenario-based 2S-SUC Formulation}
Directly calculating the expected value, $\mathbb{E}_{\bm{\xi}}[Q(\bm{x}, \bm{\xi})]$, is generally intractable due to the continuous nature of the uncertainty space. To overcome this, the uncertainty is approximated using a finite set of scenarios, resulting in a deterministic equivalent formulation known as the extensive form \cite{entensive-form}.

Let $\mathcal{S} = \{\bm{\xi}_{1}, \ldots, \bm{\xi}_{S} \}$ denote the set of $S$ representative scenarios, which are sampled from the probability distribution $\mathbb{P}$, with associated probabilities $\pi_{s}$ for each $s \in \mathcal{S}$. The second-stage problem is replicated for each scenario, and the expected recourse cost is expressed as a weighted sum over all scenarios. The resulting extensive form of the 2S-SUC model is given by:
\begin{align}
    \begin{aligned}
        \min_{\bm{x}, \{\bm{y}_{s}\}}~~& \bm{c}^{\mathsf{T}} \bm{x} + \sum_{s \in \mathcal{S}} \pi_s \, \bm{q}^{\mathsf{T}} \bm{y}_s \\
        \text{s.t.}~& \bm{x} \in \mathcal{X}, \quad \bm{y}_s \in \mathcal{Y}(\bm{x}, \bm{\xi}_s), \quad \forall s \in \mathcal{S},
    \end{aligned}
    \label{extensive-form}
\end{align}
where $\bm{y}_s$ denotes the scenario-specific recourse variables, representing generator dispatch decisions under scenario $\bm{\xi}_s$. The scenario-dependent feasible sets $\mathcal{Y}(\bm{x}, \bm{\xi}_s)$, encoding operational constraints of the UC model, are enforced independently for each scenario.

It is important to note that the number of variables and constraints in the formulation \eqref{extensive-form} grows linearly with the number of scenarios. As a result, when a large number of scenarios is required to adequately represent system uncertainty, the resulting 2S-SUC model becomes computationally burdensome. This scalability issue significantly limits the applicability of the 2S-SUC in real-world power system operations.

\section{Neural Two-Stage Stochastic Optimization}
\subsection{Conceptual Insight}
It can be observed that the recourse problem $Q(\bm{x}, \bm{\xi})$ in the 2S-SUC defines a mapping from the first-stage decisions $\bm{x}$ and scenario realization $\bm{\xi}$ to a scalar objective value. Since the primary goal of solving the UC problem is to determine the optimal commitment decision $\bm{x}$, the second-stage decision variables $\bm{y}$ can be omitted when evaluating UC outcomes. Based on this observation, a neural network can be employed to approximate the mapping $Q(\bm{x}, \bm{\xi})$, thereby enabling rapid evaluation of candidate decisions under various scenarios.

Furthermore, if the neural network surrogate is differentiable and suitably designed, it can be directly embedded into the optimization model. This facilitates an integrated solution approach to a surrogate problem that captures the effects of a large number of scenarios without explicitly modeling them. Given that the output of the neural network is a scalar, its architecture can remain structurally simple and lightweight, making its integration into mathematical programming formulations computationally feasible.

In summary, the neural stochastic model facilitates efficient solution of the 2S-SUC problem by circumventing the linear growth in problem size with respect to the number of scenarios. This approach significantly enhances computational efficiency while preserving solution quality. In the subsequent sections, we introduce the modeling architecture that enables an embedding of a neural network within a MILP formulation.

\subsection{Embedding Neural Networks into MILPs}
Mathematically, an $\ell$-layer fully-connected neural network with input $\bm{x}$ and output $\psi$ can be expressed as:
\begin{align}
       &\mathbf{h}^{1} = \sigma \left( \mathbf{W}^{1} \bm{x} + \mathbf{b}^{1} \right), 
    \\
    &\mathbf{h}^{m} = \sigma\left(\mathbf{W}^{m} \mathbf{h}^{m-1} + \mathbf{b}^{m}\right), \quad \forall m = 2, \ldots, \ell - 1, 
    \\
    &\psi = \mathbf{W}^{\ell} \mathbf{h}^{\ell - 1} + \mathbf{b}^{\ell},
\end{align}
where $\mathbf{h}^{m}$ denotes hidden states of the $m$-th hidden layer, $\mathbf{W}^{m}$ is the weight matrix of layer $m$, 
$\mathbf{b}^{m}$ is the bias vector of the $m$-th layer, and $\sigma$ is a non-linear activation function. 

A fully-connected feedforward neural network using the \texttt{ReLU} activation function, defined as $\texttt{ReLU}(a) = \max\{0, a\}$ for $a \in \mathbb{R}$, can be equivalently represented by a MILP formulation \cite{fischetti2018deep}. Specifically, for the $i$-th neuron in the $m$-th hidden layer, the activation can be written as:
\begin{equation}
    h_{i}^{m} = \texttt{ReLU} \left( \sum_{j = 1}^{N_{j}^{m}} w_{ij}^{m} h_{j}^{m-1} + b_{i}^{m} \right),
    \label{eq-relu}
\end{equation}
where $N_{j}^{m}$ is the number of inputs to layer $m$, $w_{ij}^{m}$ is the weight connecting the $j$-th neuron in layer $m-1$ to the $i$-th neuron in layer $m$, and $b_{i}^{m}$ is the corresponding bias term.

\begin{figure*}[!ht]
    \centering
    \vspace{-4mm}
    \includegraphics[width=2\columnwidth]{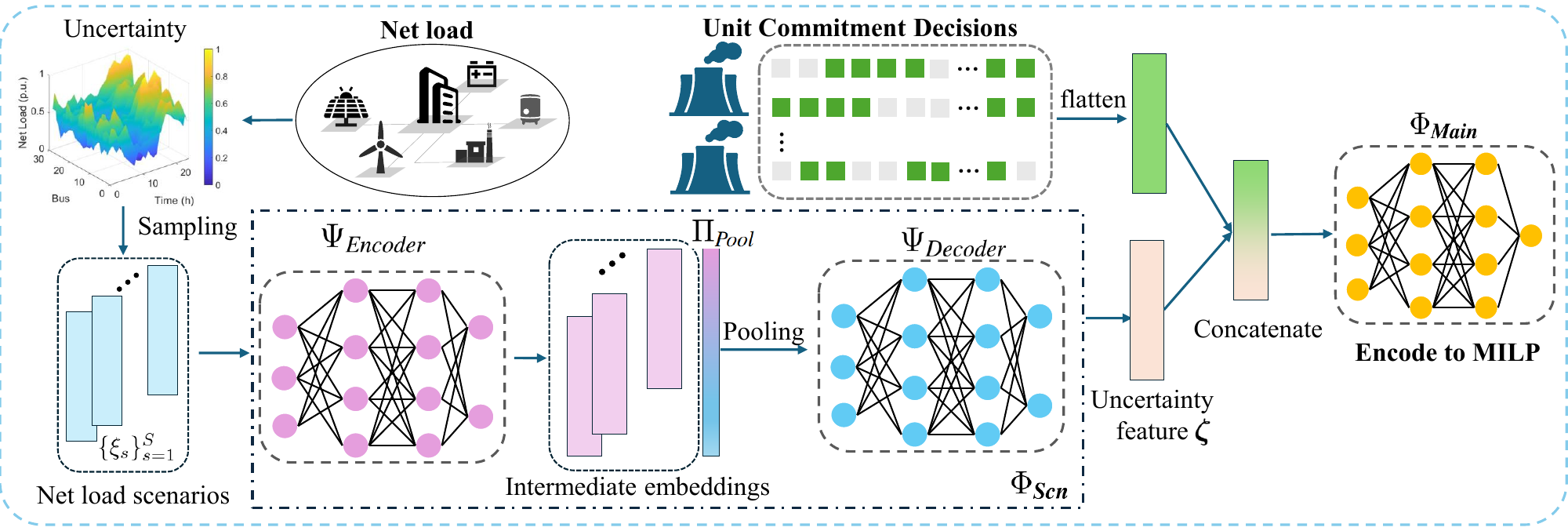}
    \vspace{-2mm}
    \caption{Architecture diagram of the proposed neural network.}
    \vspace{-4mm}
    \label{fig:nnr-architecture}
\end{figure*}

To represent the \texttt{ReLU} function within a MILP, auxiliary variables $\hat{h}_{i}^{m}$ and $\check{h}_{i}^{m}$ are introduced to capture the positive and negative components of the pre-activation value, respectively. The binary variable $z_i^m$ indicates whether the neuron is active. The \texttt{ReLU} constraint is modeled using the following MILP formulation:
\begin{subequations}
\begin{align}
    &\sum_{j = 1}^{N_{j}^{m}} (w_{ij}^{m} \hat{h}_{j}^{m-1} + b_i^{m}) = \hat{h}_i^m - \check{h}_i^m, \\
    &0 \leq \hat{h}_i^m \leq z_i^m \bar{H}_{i+}^{m}, \label{eq-relu-a} \\
    &0 \leq \check{h}_i^m \leq (1 - z_i^m) \bar{H}_{i-}^{m}, \label{eq-relu-b} \\
    &z_i^m \in \{0, 1\},
\end{align}
\label{eq-relu-milp}
\end{subequations}
where $\bar{H}_{i+}^{m}$ and $\bar{H}_{i-}^{m}$ are upper bounds on the positive and negative parts of the neuron output, respectively. Since the neural network has a nested structure and the input space is typically bounded in practice, these upper bounds (i.e., $\bar{H}_{i+}^{m}$ and $\bar{H}_{i-}^{m}$) can be estimated tightly. Using specific tight bounds rather than a large uniform constant (i.e., the big-M parameter) can improve numerical stability and reduce computational burden of the proposed surrogate model.

The correctness of the \texttt{ReLU} function is easy to verify: constraints \eqref{eq-relu-a} and \eqref{eq-relu-b}, together with the binary variable $z_i^m$, ensure that at most one of $\hat{h}_i^m$ and $\check{h}_i^m$ is non-zero, thereby correctly encoding the piecewise linear behavior of the \texttt{ReLU} activation.

\subsection{Neural Network Architecture}
This section presents the neural network designed to approximate the expected cost of the second-stage recourse problem over a set of scenarios. We refer to this model as the \emph{Neural Network Recourse Function} (NN-R). 

The NN-R model learns the mapping described in \eqref{nnr-mapping}, and its architecture is illustrated in Fig. \ref{fig:nnr-architecture}.
\begin{equation}
    \left(\bm{x}, \{ \bm{\xi}_s \}_{s=1}^S \right) \longrightarrow J = \sum_{s=1}^{S} \pi_s \, Q(\bm{x}, \bm{\xi}_s),
    \label{nnr-mapping}
\end{equation}
where the input consists of a commitment decision vector $\bm{x}$ and a set of $S$ scenarios $\{ \bm{\xi}_s \}_{s=1}^S$ sampled from the same probability distribution. The model outputs a scalar prediction $J$ representing the objective value.

The proposed neural network architecture consists of two main components:
\begin{enumerate}[leftmargin=*]
    \item[(i)] \textbf{Scenario-Embedding Network $\Phi_{\mt{Scn}}$:} 
    To capture scenario-dependent features, each scenario vector $\bm{\xi}_s$ is independently processed by a shared neural network encoder $\Psi_{\mt{Encoder}}$, yielding $S$ latent embeddings. These embeddings are then aggregated using a pooling layer $\Pi_{\mt{Pool}}$ to produce a fixed-length summary representation. The pooling layer design has great influence on the model performance. We will investigate three pooling strategies in the case study. The aggregated embedding is subsequently passed through a decoder network $\Psi_{\mt{Decoder}}$ to generate a compact scenario embedding vector, denoted as $\bm{\zeta}$. The network $\Phi_{\mt{Scn}}$ serves as a feature extractor that enables scenario compression through pooling operations, thereby facilitating generalization across varying numbers of scenarios. The decoder subsequently performs dimensionality reduction. Consequently, $\Phi_{\mt{Scn}}$ operates as a learning-based scenario reduction algorithm.

    \item[(ii)] \textbf{ReLU-Activated Main Network $\Phi_{\mt{Main}}$:} 
    The main network $\Phi_{\mt{Main}}$, implemented as a multilayer perceptron (MLP) with ReLU activation, estimates the expected second-stage cost from a joint input of scenario embedding $\bm{\zeta}$ and unit commitment decision $\bm{x}$, i.e., $\psi = \Phi_{\mt{Main}}(\bm{\zeta}, \bm{x})$. Once uncertainty feature $\bm{\zeta}$ is fixed, the MLP can be embedded into a MILP framework to optimize over $\bm{x}$ directly. Hard constraints such as the minimum up and down time constraints \eqref{UC-min-up-time} and \eqref{UC-min-down-time} can be incorporated alongside the MILP-based neural network. This ensures that the obtained UC decisions are feasible with respect to operational requirements, distinguishing the proposed method from conventional learning-based approaches that directly map scenarios to unit commitment decisions.
\end{enumerate}

The main network is implemented as a ReLU-activated MLP due to its favorable trade-off between model expressiveness and integration tractability within a MILP framework. While advanced architectures such as convolutional neural networks (CNNs), GNNs, and Transformers offer strong representational capabilities, they present substantial challenges for embedding into MILP due to their structural complexity and non-linear operations. In contrast, ReLU-activated MLPs yield piecewise linear functions, which can be efficiently encoded using standard MILP reformulations. This compatibility allows the decision variable $\bm{x}$ to be directly optimized through the embedded network while respecting operational constraints. Therefore, the use of MLP enables end-to-end decision optimization with controlled complexity, distinguishing our approach from heuristic or purely data-driven mappings.

The final NN-R model is represented as:
\begin{equation}
    \hspace{8mm} \Phi_{\mt{Main}} \left(\bm{x}, \Phi_{\mt{Scn}}\left( \{\bm{\xi}_{s}\}_{s=1}^{S} \right) \right) \approx \sum_{s=1}^{S} \pi_s Q(\bm{x}, \bm{\xi}_s).
\end{equation}
The scenario-embedding network $\Phi_{\mt{Scn}}$ can be arbitrarily complex, since only its output is used in the surrogate MILP. Although it is trained on a fixed set of $S$ scenarios, the network is capable of generalizing to arbitrary finite scenario sets. Furthermore, $\Phi_{\mt{Scn}}$ can be pre-trained independently using an autoencoder structure \cite{li2023autoencoder}, thereby reducing the data requirements compared to joint end-to-end training with the main network.

\vspace{-4mm}
\subsection{Neural Two-stage Stochastic Unit Commitment Problem}
Given the trained main network $\Phi_{\mt{Main}}^{*}$ and scenario-embedding network $\Phi_{\mt{Scn}}^{*}$, we can formulate a surrogate optimization model to represent the original 2S-SUC problem. First, the uncertainty feature vector $\bm{\zeta}^{*}$ is computed via a forward pass through the network $\Phi_{\mt{Scn}}^{*}$. Next, we simplify the first-stage decision input as $\bm{z} = \{z_{g,t}\}_{\forall g, \forall t}$, representing the binary commitment status of each generator over time. Since the start-up and shut-down indicators $u_{g,t}$ and $v_{g,t}$ can be deterministically derived from $\bm{z}$ via logic constraints, using $\bm{z}$ as the input to the main network reduces the input dimensionality and consequently lowers the computational burden during training.

The neural 2S-SUC problem is formulated as below:
\begin{subequations}
\begin{align}
    \min_{\bm{z}, \{u_{g,t}\}, \{v_{g,t}\}} & \sum_{t \in \mathcal{N}_{t}} \sum_{g \in \mathcal{N}_{g}} \left( C^{\mt{SU}}_{g} u_{g,t} + C^{\mt{SD}}_{g} v_{g,t} \right) + F(\psi) \\
    \text{s.t. \,} & \psi = \Phi_{\mt{Main}}^{*}(\bm{z}, \bm{\zeta}^{*}), \label{NN-eq} \\
               & \text{Constraints \,} \eqref{UC-logic}–\eqref{UC-min-down-time}.
\end{align}
\label{uc-surrogate-model}
\end{subequations}
where $F(\psi)$ is a rescaling function used to rescale the neural network output $\psi$ to the original cost domain, typically by inverting the min-max normalization applied during training. This rescaling function is affine and takes the follow form:
\begin{equation}
    F(\psi) = \psi \cdot (J_{\max} - J_{\min}) + J_{\min},    
\end{equation}
where $J_{\max}$ and $J_{\min}$ denote the maximum and minimum cost values used during training.

Constraint \eqref{NN-eq} denotes the network $\Phi_{\mt{Main}}^{*}$ and it is modeled using the MILP formulation shown in \eqref{eq-relu-milp}. As previously discussed, the resulting neural stochastic optimization problem is significantly easier to solve compared to the original 2S-SUC problem. In particular, by enforcing tight variable bounds \cite{ceccon2022omlt} and eliminating redundant constraints \cite{grimstad2019relu}, the solution of the linearized \texttt{ReLU} function can be very efficient, allowing the surrogate model to be solved within a few seconds \cite{anderson2020strong}. This offers a major computational advantage while still producing high-quality commitment decisions that adhere to key operational constraints.

\vspace{-4mm}
\subsection{Cold-Start and Hot-Start Modes} \label{hot-and-cold}
To effectively implement the neural stochastic optimization method for the 2S-SUC problem, several practical considerations are outlined below:

\begin{enumerate}[leftmargin=*]
    \item[(i)] \textbf{Detecting Fixed Commitment Decisions from Data:} In many real-world systems, certain units exhibit consistent commitment behavior across all training samples. For instance, nuclear power plants serving as base loads are often committed at all times. These invariant commitment decisions offer no learnable value and should be identified and excluded from the training process. Instead, such fixed commitments should be treated as hard constraints and directly enforced in the surrogate optimization model.

    \item[(ii)] \textbf{Incorporating Operational Constraints:} In addition to the minimum up and down time constraints, the surrogate model \eqref{uc-surrogate-model} can be further enhanced by incorporating additional operational constraints. For example, the constraints of power balance and unit capacity limits under the most probable scenario can be integrated to improve the quality of the resulting commitment decisions. With these additional constraints, the surrogate problem actually becomes a standard UC model augmented with neural network constraints. Although this results in a longer solution time compared to the baseline surrogate model, it yields significant improvements in solution optimality and still remains computationally efficient.

    \item[(iii)] \textbf{Solution Space Restriction:} The UC decision space grows exponentially with problem size, i.e., $2^{|\mathcal{N}_{g}||\mathcal{N}_{t}|}$, making it impractical to sample and learn across the entire solution space. However, the neural network accuracy is most critical near the optimal region. To exploit this, one can first solve a deterministic UC problem under the expected scenario to obtain a reference solution $\bm{z}^{*}$ (i.e., denoted as \emph{Kernel UC}). The surrogate optimization can be restricted to exploring solutions within a bounded Manhattan distance from the Kernel UC solution, i.e., $\|\bm{z} - \bm{z}^*\|_1 \leq \eta \cdot |\mathcal{N}_{g}||\mathcal{N}_{t}|$, where $\eta$ is a tunable parameter to control the maximum distance. Although this restriction induces heuristic behaviors, it substantially improves the solution quality and stabilizes the surrogate model by directing the optimization toward a high-potential region of the solution space.
\end{enumerate}

Note that in this paper, we refer to the basic neural stochastic optimization model \eqref{uc-surrogate-model} as the \emph{cold-start} model, whereas the enhanced surrogate model incorporating the solution space restriction strategy in (iii) is termed the \emph{hot-start} model.

\vspace{-4mm}
\subsection{Training Data Generation Strategy}
Generating high-quality training data for the neural stochastic model of 2S-SUC problem is a non-trivial task. Using naive sampling methods to generate UC decisions is unlikely to yield feasible solutions, meanwhile, relying solely on optimal 2S-SUC solutions may lead to overfitting and poor generalization due to insufficient sample diversity.

To this end, we develop an efficient strategy to generate feasible and diverse samples, involving the following steps:

\begin{enumerate}[leftmargin=*]
    \item[(i)] \textbf{Kernel UC Solution Generation:} A set of representative UC solutions, referred to as Kernel UC solutions, is first computed by solving deterministic UC problems under the most probable or extreme operational scenarios. Alternatively, the Kernel UC can be derived by directly solving the 2S-SUC with a selected subset of scenarios $S'$. The choice of $S'$ should balance computational efficiency and solution quality, as incorporating more scenarios generally improves accuracy but increases sampling time.
    
    \item[(ii)] \textbf{Perturbation and Sample Expansion:} Around each Kernel UC solution, additional samples are generated by applying structured perturbations (i.e., flipping a subset of commitment variables). This can be achieved by solving the following auxiliary optimization problem:
    \begin{subequations}
    \begin{align}
        \min_{z_{g,t}}  & \sum_{t \in \mathcal{N}_{t}} \sum_{g \in \mathcal{N}_{g}} \left( \gamma_{g,t} z_{g,t} + C^{\mt{SU}}_{g} u_{g,t} + C^{\mt{SD}}_{g} v_{g,t} \right) \\
        \text{s.t.}  & \sum_{t \in \mathcal{N}_{t}} \sum_{g \in \mathcal{N}_{g}} |z_{g,t} - z_{g,t}^{*}| \leq \epsilon \cdot |\mathcal{N}_{g}||\mathcal{N}_{t}| \\
        & \text{Constraints} \, \eqref{UC-logic}–\eqref{UC-min-down-time}
    \end{align}
    \end{subequations}
    Here, $z_{g,t}^{*}$ denotes the commitment status of the Kernel UC solution, and $\gamma_{g,t}$ are randomly generated coefficients sampled uniformly from the interval $[-1, 1]$. A positive value of $\gamma_{g,t}$ encourages the unit to be off during time interval $t$, whereas a negative value promotes it to be on. The scalar parameter $\epsilon \in (0,1)$ controls the allowable perturbation level, typically set to $\epsilon = 0.2$. This approach maintains feasibility while introducing controlled randomness to enhance dataset diversity and representativeness.
    
    \item[(iii)] \textbf{OPF-based Fast Sampling:} Once a sufficient number of UC solutions have been obtained, it is no longer necessary to solve the computationally intensive 2S-SUC problems. Instead, given a UC solution and a series of scenario realizations, optimal power flow (OPF) problems can be solved to evaluate the second-stage operational cost. The training label can be computed by a weighted summation over a designated set of scenarios: $J = \sum_{s'=1}^{S'} \pi_{s'} Q(\bm{z}^{*}, \bm{\xi}_{s'})$.
    This procedure enables efficient and scalable data generation without requiring repeated full stochastic optimization.

    \item[(iv)] \textbf{Lagrange Function-based Labeling:} Sampling efficiency can be further improved by utilizing the Lagrange function derived from the Kernel UC in step (i) to label the OPF instances obtained in step (iii). This labeling strategy offers two primary advantages. First, by incorporating the Lagrange function, the inherently complex and high-dimensional learning surface is effectively linearized, making it easier for the neural network to learn. Second, since perturbations can result in infeasible instances, the Lagrange function offers a principled way to explicitly identify and label such infeasibility.
\end{enumerate}

This strategy eliminates the need to solve numerous full-scale 2S-SUC problems and generates a diverse, feasible training dataset that enhances neural network generalization across system operating conditions.

\vspace{-3mm}
\section{Numerical Study}
\vspace{-1mm}
\subsection{Numerical Study Setup}
\vspace{-1mm}
To validate the effectiveness of the proposed method, three standard test systems from MATPOWER \cite{matpower}, including the 5-bus, 30-bus, and 118-bus systems, are employed for comparison against state-of-the-art benchmark approaches. The neural network surrogate models are trained using PyTorch, and the resulting surrogate optimization problems are solved using the OMLT toolbox \cite{ceccon2022omlt} in conjunction with Pyomo \cite{hart2011pyomo}. The OMLT framework integrates big-M acceleration techniques and dead neuron detection strategies, which achieves substantial computational speedup. The OMLT can efficiently solve optimization problems that include neural networks with up to one thousand \texttt{ReLU} activation functions in just a few seconds. The neural network architectures adopted for the 5-bus and 30-bus systems consist of two hidden layers with dimensions $64$–$64$–$1$, whereas the 118-bus system employs a deeper architecture comprising three hidden layers with dimensions $256$–$128$–$64$–$1$. Detailed hyperparameter setups are provided in Table~\ref{table-nn-parameters}. The hyperparameters $\eta$ and $\epsilon$ are both set to 0.2. All experiments are conducted using Gurobi 11 on a workstation equipped with an Intel i9-9900X @ 3.5\,GHz processor and 64\,GB of RAM. 

Two state-of-the-art baseline methods are considered:
\begin{enumerate}
    \item[(1)] \textbf{Gurobi}, which directly solves the 2S-SUC problem using the extensive form in MILP formulation.
    \item[(2)] \textbf{CCG}, the column-and-constraint generation method \cite{zeng2013ccg}, originally developed for two-stage robust optimization, which can be effectively and efficiently applied to solve the 2S-SUC problem in an iterative manner.
\end{enumerate}

The system uncertainty is modeled as net load variation, where the net load at each bus is independently and uniformly sampled within the range of 70\% to 100\% of its maximum demand.

\begin{table}[tbp]
\centering
\vspace{-3mm}
\caption{Hyperparameter Settings for Systems of Different Sizes}
\vspace{-1mm}
\label{table-nn-parameters}
\begin{tabularx}{\linewidth}{lXX}
\toprule
\textbf{Parameter} & \textbf{5-Bus / 30-Bus} & \textbf{118-Bus} \\
\midrule
Batch size & 32 & 64 \\
Learning rate & $10^{-3}$ & $10^{-4}$ \\
L1 weight penalty & $10^{-4}$ & $2 \times 10^{-4}$ \\
L2 weight penalty & $10^{-5}$ & $2 \times 10^{-5}$ \\
Optimizer & \{Adam, Adagrad\} & Adam \\
Dropout & 0.01 & 0.01 \\
\# Epochs & 200 & 500 \\
ReLU hidden dimension & \{64, 64\} & \{256, 128, 64\} \\
Encoder hidden dimension & \{64, 24\} & \{64, 64, 24\} \\
Decoder hidden dimension & \{64, 32\} & \{128, 64\} \\
\bottomrule
\end{tabularx}
\vspace{-6mm}
\end{table}



\vspace{-4mm}
\subsection{Evaluation of Optimality and Computation Efficiency}
\vspace{-1mm}

\begin{table*}[!t]
\vspace{-4mm}
\renewcommand{\arraystretch}{1.2}
\caption{Performance Comparison of Different Methods on Solving Two-Stage Stochastic UC Problems}
\vspace{-1mm}
\label{tab-comparison}
\begin{tabularx}{\textwidth}{llYYYYYYYYYY}
\toprule
\multirow{2}{*}{\vspace{-2mm}\textbf{System}} & \multirow{2}{*}{\vspace{-2mm}\textbf{\#Scen.}} 
& \multicolumn{3}{c}{\textbf{Avg. Objective (\$)}} 
& \multicolumn{2}{c}{\textbf{Avg. Gap (\%)}} 
& \multicolumn{3}{c}{\textbf{Solution Time (s)}} 
& \multicolumn{2}{c}{\textbf{Speed Up}} \\
\cmidrule(lr){3-5} \cmidrule(lr){6-7} \cmidrule(lr){8-10} \cmidrule(lr){11-12}
& & \textbf{Gurobi} & \textbf{CCG} & \textbf{Proposed} 
& \textbf{CCG} & \textbf{Proposed} 
& \textbf{Gurobi} & \textbf{CCG} & \textbf{Proposed} 
& \textbf{CCG} & \textbf{Proposed} \\
\midrule
\multirow{3}{*}{5-bus} 
& $S$ = 10  & 205{,}891 & 205{,}932 & 205{,}941 & 0.02 & 0.02 & 0.5   & 1.1   & \textbf{0.06} & 0.5$\times$ & \textbf{8.3$\times$} \\
& $S$ = 50  & 204{,}316 & 204{,}439 & 204{,}479 & 0.06 & 0.06 & 4.2   & 1.8   & \textbf{0.08} & 2.3$\times$ & \textbf{52.5$\times$} \\
& $S$ = 100 & 202{,}673 & 202{,}734 & 202{,}916 & 0.03 & 0.05 & 15.6  & 3.8   & \textbf{0.08} & 4.1$\times$ & \textbf{195.0$\times$} \\
\midrule
\multirow{3}{*}{30-bus} 
& $S$ = 10  & 779{,}091 & 779{,}714 & 781{,}350 & 0.05 & 0.29 & 6.2   & 3.2   & \textbf{0.28} & 1.9$\times$ & \textbf{22.1$\times$} \\
& $S$ = 50  & 767{,}383 & 768{,}304 & 769{,}992 & 0.03 & 0.34 & 48.8  & 19.5  & \textbf{0.31} & 2.5$\times$ & \textbf{157.4$\times$} \\
& $S$ = 100 & 760{,}746 & 761{,}887 & 763{,}637 & 0.06 & 0.38 & 100.5 & 32.1  & \textbf{0.38} & 3.1$\times$ & \textbf{264.5$\times$} \\
\midrule
\multirow{3}{*}{118-bus} 
& $S$ = 10  & 1{,}657{,}120 & 1{,}658{,}943 & 1{,}666{,}400 & 0.03 & 0.56 & 215.7  & 76.8   & \textbf{16.5}  & 2.8$\times$ & \textbf{13.1$\times$} \\
& $S$ = 50  & 1{,}622{,}308 & 1{,}625{,}066 & 1{,}632{,}366 & 0.06 & 0.62 & 1782.5 & 574.5  & \textbf{17.8}  & 3.1$\times$ & \textbf{100.1$\times$} \\
& $S$ = 100 & 1{,}615{,}052 & 1{,}618{,}444 & 1{,}630{,}556 & 0.07 & 0.96 & 4340.8 & 1349.8 & \textbf{18.6}  & 3.2$\times$ & \textbf{233.4$\times$} \\
\midrule
\textbf{Average} & -- & 868{,}287 & 869{,}496 & 873{,}070 & 0.05 & \textbf{0.36} & 723.87 & 229.18 & \textbf{6.01} & 2.62$\times$ & \textbf{116.3$\times$} \\
\bottomrule
\end{tabularx}
\vspace{-5mm}
\end{table*}

The comparison results are summarized in Table \ref{tab-comparison}, where each metric is evaluated using 100 test instances. The column labeled \textbf{\#Scen} denotes the number of scenarios contained in a 2S-SUC instance, which reflects the problem scale. Generally, larger values of \#Scen correspond to higher model complexity. The optimality \textbf{gap} measures the solution quality, which is calculated using the Gurobi solution as the benchmark.

As shown in Table \ref{tab-comparison}, in terms of solution quality, the proposed method yields objective values that are reasonably close to those obtained by the Gurobi solver, with gap consistently below $1\%$ across all test cases. For example, in the 118-bus system with 100 scenarios, the gap is 0.96\%, indicating that despite the approximation introduced by the surrogate, it maintains sufficient accuracy for the solutions of SUC problems. 

Regarding computational efficiency, the proposed method offers a notable reduction in solution time compared to both the Gurobi and CCG algorithm. This is primarily due to the use of pre-trained neural network surrogates, which shift the scenario-dependent complexity to the offline phase. In large-scale instances, such as the 118-bus system with 100 scenarios, the proposed method solves the problem in under 20 seconds, whereas Gurobi and CCG require significantly longer runtimes. Compared to CCG, which relies on iterative decomposition into master and subproblems, the proposed method formulates a single MILP using a surrogate model, thereby avoiding the need for any decomposition. Nonetheless, across the evaluated test cases, the trade-off between speed and optimality appears favorable, particularly when rapid computation is prioritized.

A sensitivity analysis with respect to the number of scenarios further illustrates the computational efficiency of the proposed method. Unlike conventional approaches such as Gurobi or CCG, where the problem size grows proportionally with the number of scenarios, the size of the surrogate-based MILP remains constant regardless of the scenario count. This is because the trained neural network encapsulates the scenario-dependent second-stage cost function. The MILP reformulation depends solely on the network architecture, and it is not related to the number of input samples used during training or evaluation.

However, increasing the number of scenarios can lead to more complex uncertainty features being evaluated through the surrogate, which in turn may activate a greater number of ReLU neurons in the network. As the surrogate optimization model includes constraints for active neurons, a higher activation density can marginally increase the complexity of the resulting MILP. This may explain the slight growth in solution time observed in the proposed method as the scenario count increases from 10 to 100, despite the overall problem structure remaining fixed. For example, in the 30-bus system, the solution time of the proposed method increases from 0.29 seconds at $S=10$ to 0.38 seconds at $S=100$, whereas the corresponding runtimes for Gurobi increase from 6.2 to 100.5 seconds. This contrast highlights that while the surrogate-based formulation is not entirely immune to scenario-induced complexity, its sensitivity is substantially lower than that of traditional solvers. Importantly, the observed runtime growth remains modest and predictable, suggesting that the method retains its computational advantage even under high-resolution uncertainty representations.

\vspace{-5mm}
\subsection{Computation Efficiency and Optimality Trade-Off}

While the proposed method demonstrates notable improvements in computational efficiency, it exhibits a larger optimality gap compared to the Gurobi solver. To better quantify the improvement in computation efficiency and further examine the trade-off between solution quality and computational time, two additional experiments are conducted. First, the time required for Gurobi to reach the same optimality gap as that produced by the proposed method is measured, thereby quantifying the corresponding time savings; this test is referred to as \textbf{Gurobi-to-NNS-gap}. Second, the solution obtained from the proposed method is used as a warm-start for the Gurobi solver, and the time required for Gurobi to refine this solution to optimality is evaluated; this test is referred to as \textbf{NNS-to-optimal}, where \textbf{NNS} refers to the neural network surrogate approach. These analyses help quantify the improvement in computation efficiency achieved by the proposed algorithm when we set a fixed target on solution quality.

\begin{figure}[tbp]
    \centering
    \vspace{-4mm}\includegraphics[width=6.5cm]{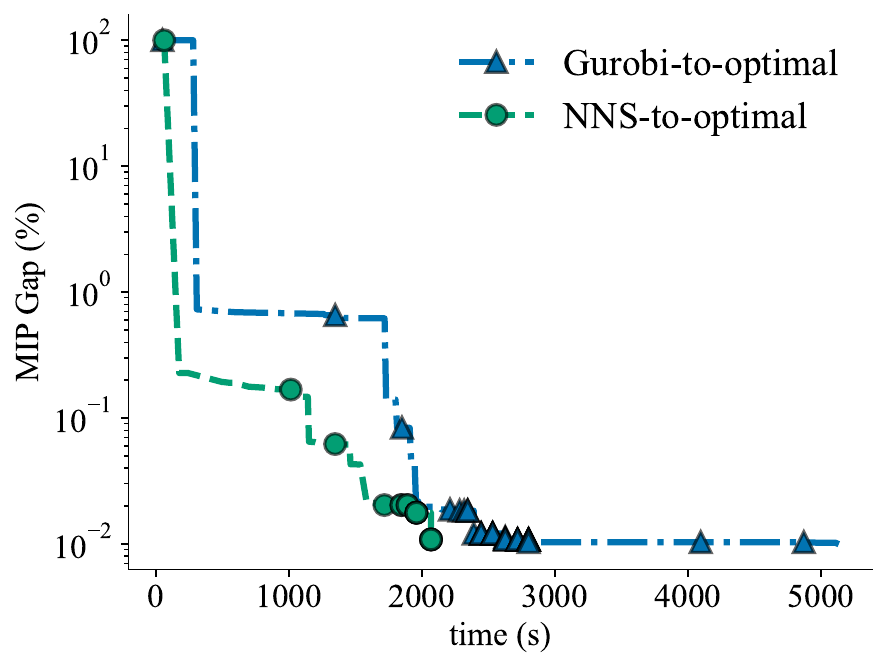}
    \vspace{-4mm}
    \caption{Optimality gap evolution over time for the 118-bus system with 100 stochastic scenarios.}
    \vspace{-2mm}
    \label{fig-time-gap}
\end{figure}

The 118-bus system under 10, 50, and 100 scenarios is used for testing. The test results are summarized in Table~\ref{tab:new-performance}. The Fig.~\ref{fig-time-gap} shows the solution quality as a function of time for Gurobi-to-optimal and NNS-to-optimal under the 100-scenarios settings. 
These two comparisons are especially useful for quantifying computational efficiency gains. 

First, a comparison between NNS and Gurobi-to-NNS-gap reflects the time required to achieve the same solution quality. Under 100 scenarios, the NNS method obtains a solution in 18.3 seconds, whereas Gurobi requires 1724.3 seconds to reach a solution of comparable quality. This represents nearly a 100-fold reduction in computational time, demonstrating substantial efficiency gains when moderate optimality gaps are acceptable. A similar trend is observed for 50 scenarios, where the NNS method achieves a solution in 18.2 seconds compared to 255.4 seconds for Gurobi. However, when the number of scenarios is reduced to 10, the advantage disappears, with NNS requiring 16.9 seconds—slightly more than Gurobi's 14.5 seconds. 

Second, the comparison between NNS-to-optimal and Gurobi-to-optimal assesses the effectiveness of using NNS as a warm-start for solving for full optimality. With 100 scenarios, reaching MIP gap of 0.01\% from the NNS solution takes 2069.9 seconds, compared to 5132.6 seconds using Gurobi without warm-start - yielding a nearly 60\% reduction in computation time, as shown in Fig.~\ref{fig-time-gap}. At 50 scenarios, the computing time from NNS solution to desired optimality level is 766.0 seconds versus 1533.5 seconds for Gurobi. Even with just 10 scenarios, the warm-start strategy reduces the solution time from 154.4 to 55.6 seconds, achieving a relative improvement of approximately 64\%. These results indicate that while absolute time savings diminish in lower-dimensional settings, the relative efficiency gain remains significant.

\begin{table}[t]
\centering
\vspace{-2mm}
\caption{Solution Time for Different Setings on the 118-Bus System}
\vspace{-1mm}
\label{tab:new-performance}
\renewcommand{\arraystretch}{1.2}
\begin{tabularx}{0.95\linewidth}{l>{\centering\arraybackslash}X>{\centering\arraybackslash}X>{\centering\arraybackslash}X}
\toprule
\multirow{2}{*}{\vspace{-2mm}\textbf{Solution Time (s)$^{\dagger}$}} & \multicolumn{3}{c}{\textbf{Number of Scenarios}} \\
\cmidrule(lr){2-4}
& 10 & 50 & 100 \\
\midrule
Gurobi-to-NNS-gap      & 14.5 & 255.4 & 1724.3 \\
NNS                    & 16.9 & 18.2 & 18.3 \\
\midrule
Gurobi-to-optimal      & 154.4 & 1533.5 & 5132.6 \\
NNS-to-optimal         & 55.6 & 766.0 & 2069.9 \\
\bottomrule
\end{tabularx}
\\[0.5em]
\raggedright \textit{~~~Notes:} \textit{$^{\dagger}$The MIP Gap for the Gurobi solver is set to 0.01\%} 
\vspace{-5mm}
\end{table}

\begin{figure}[!tbp]
    \centering
    \includegraphics[width=2in]{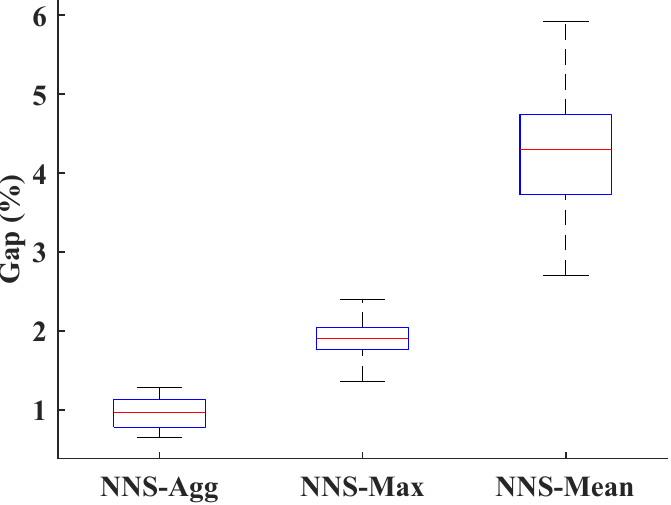}
    \vspace{-2mm}
    \caption{The MIP gap distribution of the 118-bus system under 100 stochastic scenarios across different pooling strategies.}
    \vspace{-6mm}
    \label{fig-error-boxplot}
\end{figure}

\vspace{-4mm}
\subsection{Impact of Pooling Strategy on Error Distribution}
To further evaluate the robustness of the proposed method, this section examines the distribution of solution errors under different uncertainty pooling strategies. Specifically, three pooling mechanisms are investigated:
\begin{enumerate}
    \item[(i)] \textbf{NNS-Max}: applies max pooling to the encoded scenario features;
    \item[(ii)] \textbf{NNS-Mean}: applies mean pooling to the encoded scenario features;
    \item[(iii)] \textbf{NNS-Agg}: combines both max and mean pooled features.
\end{enumerate}

The error distribution for the 118-bus system with 100 stochastic scenarios is illustrated in Fig. \ref{fig-error-boxplot}. As shown in the figure, the \textbf{NNS-Agg} method, which integrates both maximum and mean pooling features, achieves the best performance, with a median, maximum, and minimum MIP gap of 0.966\%, 1.282\%, and 0.645\% respectively. For the \textbf{NNS-Max} method, the median, maximum, and minimum MIP gaps are 1.909\%, 2.491\%, and 1.363\%, respectively. In contrast, the \textbf{NNS-Mean} method yields a median, maximum, and minimum MIP gap of 4.297\%, 5.919\%, and 2.702\%. These results indicate that the aggregation pooling strategy outperforms the individual pooling methods, while mean pooling exhibits the worst performance. 

\vspace{-4mm}
\subsection{Scalability under Hot‑Start and Cold‑Start Modes}

\begin{table}[t]
\centering
\vspace{-4mm}
\caption{Scalability Performance under Different Scenario Sizes on the 118-Bus System}
\vspace{-2mm}
\label{tab-scalability}
\renewcommand{\arraystretch}{1.2}
\resizebox{\linewidth}{!}{%
\begin{tabularx}{1.1\linewidth}{llllll}
\toprule
\multirow{2}{*}{\vspace{-2mm}\textbf{Method}} & \multirow{2}{*}{\vspace{-2mm}\textbf{Metric}} & \multicolumn{4}{c}{\textbf{Number of Scenarios}} \\
\cmidrule(ll){3-6}
& & 100 & 200 & 500 & 1000 \\
\midrule
\multirow{2}{*}{Gurobi}    
& Objective (\$)   & 1{,}615{,}052 & 1{,}614{,}822 & 1{,}614{,}218 & 1{,}613{,}558 \\
& Time (s)         & 4{,}340 & 8{,}632 & 22{,}562 & 46{,}310 \\
\midrule
\multirow{3}{*}{CCG}       
& MIP Gap (\%)         & 0.072 & 0.056 & 0.058 & 0.063 \\
& Time (s)         & 1{,}350 & 2{,}256 & 5{,}232 & 9{,}213 \\
& Speedup          & \textbf{3.21$\times$} & \textbf{3.82$\times$} & \textbf{4.31$\times$} & \textbf{5.02$\times$} \\
\midrule
\multirow{3}{*}{Cold-start} 
& MIP Gap (\%)         & 0.96 & 1.32 & 1.85 & 1.92 \\
& Time (s)         & 18.6 & 18.7 & 19.5 & 20.6 \\
& Speedup          & \textbf{233.4$\times$} & \textbf{461.6$\times$} & \textbf{1157.0$\times$} & \textbf{2248.1$\times$} \\
\midrule
\multirow{5}{*}{Hot-start} 
& MIP Gap (\%)              & 0.56 & 0.58 & 0.62 & 0.66 \\
& Time-S$^{\dagger}$ (s) & 25.2 & 25.8 & 26.5 & 27.6 \\
& Time-H$^{\ddagger}$ (s) & 10.3 & 26.1 & 103.8 & 241.0 \\
& Tol. Time (s)          & 35.5 & 51.9 & 130.3 & 268.6 \\
& Speedup                & \textbf{122.3$\times$} & \textbf{166.3$\times$} & \textbf{173.2$\times$} & \textbf{172.4$\times$} \\
\bottomrule
\end{tabularx}%
}
\\[0.5em]
\raggedright \textit{Notes:} \\
\textit{$\dagger$ Time-S denotes the solution time of the surrogate MILP formulation using a hot-start initial point with $\eta=0.2$.} \\
\textit{$\ddagger$ Time-H represents the time required to obtain a hot-start UC point, derived by solving a 2SO-UC with relaxed binary variables and without transmission constraints.}
\vspace{-4mm}
\end{table}

As introduced in Section \ref{hot-and-cold}, we consider two initialization strategies for the proposed surrogate method: \textit{cold-start} and \textit{hot-start} modes. The surrogate neural network in cold-start mode solves the problem without any prior solution information. In contrast, the hot-start mode incorporates a preliminary solution to restrict the search space of the surrogate MILP, potentially improving accuracy at the cost of increased model complexity. Additionally, it is mentioned that the results in the Table \ref{tab-comparison} are obtained under cold-start mode.

Table \ref{tab-scalability} presents the scalability performance of the proposed method under both modes on the IEEE 118-bus system with 100, 200, 500, and 1,000 scenarios. For the cold-start mode, the solution time remains relatively stable around 19--21 seconds. This is because the surrogate MILP formulation size is independent of the number of scenarios. The variation in the number of scenarios influences only the uncertainty features input to the neural network. As the number of scenarios increases, the greater diversity in input features may activate slightly more neurons, leading to a marginal increase in solution time. In terms of optimality, the MIP gap ranges from 0.96\% to 1.92\%. The trend of increasing error appears to plateau, indicating that beyond a certain point, additional scenarios offers diminishing returns to the UC solution by providing limited incremental value in representing system uncertainty. Notably, the cold-start approach achieves substantial speedup, ranging from \textbf{233.4$\times$} to \textbf{2248.1$\times$}, highlighting its computational efficiency, particularly in large-scale instances.

For the hot-start mode, the initial solution is obtained by solving a relaxed version of the 2S-SUC problem (with binary variables relaxed and transmission constraints omitted), referred to as Time-H. The surrogate MILP then solves the problem in a local region (i.e., with $\eta = 0.2$) around the initial solution, with the corresponding runtime recorded as Time-S. While the solution time of the surrogate MILP increases to approximately 26 seconds due to the additional constraints associated with hot-start guidance, the total computational cost (i.e., Time-H + Time-S) remains significantly lower than that of solving the full 2S-SUC. The total runtime leads to speedup factors between \textbf{122.3$\times$} and \textbf{172.4$\times$}, which are lower than the cold-start case, but are still notable. Importantly, the optimality of solutions obtained in the hot-start mode improves, with MIP gap reduced to 0.56\%--0.66\%, indicating the effectiveness of leveraging a high-quality initial solution point.

Finally, the performance of the CCG method is also presented as a decomposition-based benchmark. CCG exhibits strong solution quality with an average MIP gap of 0.062\%. The speedup improves with the number of scenarios, increasing from \textbf{3.21$\times$} to \textbf{5.02$\times$}, demonstrating better scalability than the extensive form approach. However, even with such improvements, CCG remains approximately two orders of magnitude slower than the proposed surrogate optimization method. Overall, the proposed approach achieves favorable trade-offs between solution quality and computational efficiency, making it suitable for large-scale stochastic UC applications.

\vspace{-2mm}
\section{Conclusion and Discussion}
This paper proposes a neural two-stage stochastic optimization approach for accelerating the solution of large-scale 2S-SUC problems. By embedding a trained ReLU-activated neural network into a MILP formulation, the proposed method approximates the second-stage recourse cost and integrates it directly into the first-stage decision-making process. A scenario-embedding network is introduced to enable flexible and scalable handling of uncertainty features, thereby supporting data-driven scenario reduction. 

Numerical studies on IEEE benchmark systems confirm that the proposed method achieves near-optimal solutions with a MIP gap below 1\%, while significantly reducing computational time compared to conventional extensive form and decomposition-based approaches. Furthermore, the surrogate model's computational complexity is independent of the number of scenarios, providing strong scalability for large-scale applications. Both cold-start and hot-start modes are evaluated, and results show that the hot-start model further improves optimality with minimal additional runtime. Overall, the proposed surrogate optimization framework offers an effective, scalable, and operationally feasible solution to the 2S-SUC problem, advancing practical implementation of stochastic optimization in real-world power system operations. 

While this work makes substantial progress in accelerating the solution of the 2S-SUC problem, extending the proposed approach to large-scale practical systems remains an important area for future research. In systems with over 10,000 generating units, The corresponding neural network may require more than 20,000 neurons, presenting significant challenges for state-of-the-art MILP solvers. To address this, incorporating a suitable encoder to reduce the dimensionality of unit commitment decisions is both necessary and a promising research direction.

\vspace{-4mm}


\bibliographystyle{IEEEtran}
\bibliography{Refs}
\end{document}